\documentclass[11pt]{article}
\usepackage[dvips]{graphicx}
\begin{document}

\begin{center}
\textbf{Derivation of Hodgkin-Huxley equations for a Na+  channel from a master equation for coupled activation and inactivation}
 \end{center}
\begin{center} S. R. Vaccaro  \end{center}
\begin{center} 
{\em Department of Physics, University of Adelaide, Adelaide, South Australia, 
5005, 
Australia} \\
  \end{center}
{\em svaccaro@physics.adelaide.edu.au} \\

{\bf Abstract}

\begin{quotation}
The Na+ current in nerve and muscle membranes may be described in terms of the activation variable m(t) and the
 inactivation variable h(t), which are dependent on the transitions of S4 sensors of each of the Na+ channel domains
 DI to DIV. The time-dependence of the Na+ current and the rate equations satisfied by  m(t)  and h(t) may be derived 
from the solution to a master equation which describes the coupling between two or three activation sensors regulating
 the Na+ channel conductance and a two stage inactivation process. If the inactivation rate from the closed or open states
 increases as the S4 sensors activate, a more general form for the Hodgkin-Huxley expression for the open state
 probability may be derived where m(t) is dependent on both activation  and inactivation processes. The voltage
 dependence of  the rate functions  for inactivation and recovery from inactivation are consistent with the empirically 
determined  expressions, and exhibit saturation for both depolarized and hyperpolarized clamp potentials.
\end{quotation}
\newpage

{\bf INTRODUCTION}  

 The opening and subsequent inactivation of Na+ channels and the activation of K+ channels generate
 the action potential in nerve and muscle membranes \cite{hh}. The time-dependence of the Na+ current in the squid axon may 
be described in terms of the expression $m(t)^3 h(t)$ where the activation variable m(t) and inactivation variable h(t) satisfy 
the rate equations
\begin{equation}
\frac{dm}{dt}= \alpha_{m}-(\alpha_{m}+\beta_{m})m,  \label{ratem}
\end{equation}

\begin{eqnarray}
\frac{dh}{dt}= \alpha_{h}-(\alpha_{h}+\beta_{h})h,  \label{rateh}
\end{eqnarray}
and $\alpha_{m}$, $\beta_{m}$, $\alpha_{h}$, and $\beta_{h}$ are voltage dependent rate functions for activation and inactivation 
transitions across the membrane.

    The Hodgkin-Huxley (HH) description of the Na+ current is equivalent to an 8-state master equation where three independent  
voltage sensors may activate and open the channel and independent inactivation may occur from each of the closed or open states 
 \cite{arm, hille,keener}. Although this master equation is not consistent with the measurement of an almost zero Na+ current during 
 repolarization  of an inactivated channel, by assuming that the deinactivation rate to the open state is  zero
 but the rate to closed states increases as the S4 sensors deactivate  \cite{kb}, and that the rate functions satisfy microscopic  reversibility,
 the model provides a good description of the recovery from inactivation,  and the Na+ current during a depolarizing clamp  \cite{cgabc}.

In this paper, it is shown  that the  Hodgkin-Huxley expression for the Na+ current and the rate equations for activation and inactivation
 may be derived from a master  equation, which describes the coupling between two or three activation sensors  regulating the 
 Na+ channel  conductance and a two-stage inactivation process. For a Na+ channel with two activation sensors,   where the deactivation 
rate during repolarization is slower between inactivated states than  between closed or open states,  only four of the terms  of the
solution to a six state master  equation contribute to the dynamics, and if the activation sensors are independent, the
open state $O(t)$  may be expressed as $m(t)^2 h(t)$. The voltage dependence of  the rate functions  for inactivation and recovery
from inactivation have a similar form to empirical expressions for Na+ channels 
 \cite{hh,kb}, and in particular, the exponential variation exhibits saturation for both depolarized  and hyperpolarized clamp potentials.

\newpage

{\bf  VOLTAGE CLAMP OF A Na+ CHANNEL WITH TWO ACTIVATION SENSORS}

The Na+ channel protein is comprised of four domains DI to DIV, each containing  an S4 segment with positively charged residues
 at every third position \cite{hille}. Based on voltage clamp fluorometry, it has been shown that, in response to membrane
 depolarization, the transverse motion of the charged S4 segments of the Na+ channel domains DI to DIII is associated with
 activation, whereas the slower movement of DIV S4 is correlated with the binding of  an intracellular hydrophobic motif  that 
blocks the flow of ions through the inner mouth of the pore  \cite{cb}. This may occur for small depolarizations when the ion
 channel is usually closed (closed-state inactivation) or for larger depolarizations when the S4 segments of the domains D1 to D3
 are activated (open-state inactivation). 

However,  during repolarization of an inactivated Na+ channel, the OFF gating charge has a fast component which may be 
attributed to the motion of the DI and DII S4 segments, and a slow component,   the "immobilized" portion, that is generated
by the conformational changes of the DIII and DIV S4 segments  \cite{ab,crgfb}.  For an  inactivation modified 
mutant of the human heart Na+ channel, it has been estimated that the DIV S4 sensor contributes approximately 30 \% to the OFF 
charge, and approximately 20 \% may be attributed to the DIII S4 sensor which is only immobilized when the inactivation gate is intact. 
The slow component of the OFF gating charge  has the same time-course as the Na+ channel recovery from inactivation, and 
therefore, the rate-limiting step is the motion of the DIV S4 sensor and not the unbinding of the inactivation gate \cite{shk}.  

In order to account for the effect of double-cysteine mutants of S4 gating charges on the ionic current of the bacterial Na+ channel
 NaChBac, it has been proposed that at least two transitions are required during the activation of each voltage sensor   \cite{dystc}. 
 This conclusion is consistent with an earlier result that cross-linking a DIV S4 segment from the extracellular surface inhibits
 inactivation during membrane depolarization whereas cross-linking the same segment from the inside inhibits activation of the 
Na+ channel, and therefore, the DIV S4 sensor translocates across the membrane in two stages   \cite{hdb,arm1}. 
The measurement of  currents for charge neutralized segments in each domain of the Na+ channel gives additional
 support to the conclusion that the two-stage activation of the DIV S4 sensor is correlated with ion channel inactivation \cite{cgabc}.

  In this section, we assume that the activation of two voltage sensors  regulating the  Na+ channel  conductance
 (DIII S4 and the S4 segment  of either the DI or DII domains)  is coupled to a two-stage inactivation process 
 (see Fig. 1), and therefore, the kinetics may be described by a master equation where the occupation probabilities
 of the closed states $C_1$, $C_2$, $A_1$ and $A_2$,  the open 
states $O$ and $A_3$, and  the inactivated (or blocked) states $B_1$, $B_2$ and $B_3$ are determined by
\begin{eqnarray}
\frac{dC_1}{dt} & = &  -(\alpha_{i1} + \alpha_C)C_1(t)  + \beta_C C_2(t) + \beta_{i1} A_1(t) \label{9c1} \\
\frac{dC_2}{dt} & = &  -(\alpha_{i2} + \alpha_O + \beta_C)C_2(t)  + \alpha_C C_1(t) + \nonumber  \\
                &   &  \beta_O O(t) + \beta_{i2} A_2(t) \label{9c2} \\
\frac{dO}{dt} & = &  \alpha_O C_2(t) - (\beta_O + \alpha_{i3})O(t) + \beta_{i3} A_3(t) \label{9o} \\
\frac{dA_1}{dt} & = &  \alpha_{i1} C_1(t) -(\alpha_{A1} + \beta_{i1} + \gamma_{i1})A_1(t)  \nonumber \\
               &   & + \delta_{i1} B_1(t) + \beta_{A1} A_2(t) \label{9a1} \\
\frac{dA_2}{dt} & = & \alpha_{i2} C_2(t) -(\alpha_{A2} + \beta_{A1} + \beta_{i2} + \gamma_{i2})A_2(t)  \nonumber \\
                &   & +\delta_{i2}B_2(t) + \alpha_{A1} A_1(t) + \beta_{A2} A_3(t) \label{9a2} \\
\frac{dA_3}{dt} & = & \alpha_{i3} O(t) -(\beta_{A2} + \beta_{i3} + \gamma_{i3})A_3(t)  \nonumber \\
                &   & +\delta_{i3}B_3(t) + \alpha_{A2} A_2(t)  \label{9a3} \\
\frac{dB_1}{dt} & = & \gamma_{i1} A_1(t)-(\alpha_{B1} + \delta_{i1}) B_1(t) +\beta_{B1} B_2(t) \label{9b1} \\
\frac{dB_2}{dt} & = & \gamma_{i2} A_2(t) +\alpha_{B1} B_1(t) +\beta_{B2} B_3(t) - \nonumber  \\
                &   & (\alpha_{B2} + \beta_{B1} + \delta_{i2}) B_2(t), \label{9b2} \\
\frac{dB_3}{dt} & = & \gamma_{i3} A_3(t) +\alpha_{B2} B_2(t)  - ( \beta_{B2} + \delta_{i3}) B_3(t). \label{9b3}
\end{eqnarray}

The master equation may be derived from a Smoluchowski equation applied to the resting and barrier
 regions of an energy landscape for each of the S4 sensors in the domains DI to DIV \cite{vac1,vac2}.
 The translocation of the S4 segment through the gating pore for Na+ (or K+)  channels requires sufficient 
energy to overcome several barriers that are dependent on the Coulomb force between positively charged
 residues on the S4 sensor and negatively charged residues on neighboring helices, the dielectric boundary
 force, the electric field between internal and external aqueous crevices, and hydrophobic forces \cite{llg}.
 It is assumed that the transition rates for each stage of inactivation are dependent on single barrier activation,
 and therefore, are proportional to exp(-U) where  U is the voltage dependent height of the barrier \cite{kram}.
However, if the Na+ channel S4 sensors of the DI, DII or DIII domains are activated in two stages, the rate 
functions  $\alpha_m$ and  $\beta_m$ may be approximated by two-state expressions  \cite{vac3}. 

    In order to simplify the solution of Eqs. (\ref{9c1}) to (\ref{9b3}), it is initially assumed that

 \begin{equation}
\alpha_{ik} = \alpha_{i1}, \beta_{ik} = \beta_{i1}, \gamma_{ik} = \gamma_{i1},  \label{alfi}
\end{equation} 
for each $k$, and to ensure that the Na+ current recovers from inactivation when the S4 sensors that regulate Na+ conductance deactivate, it is further assumed that
\begin{equation}
\delta_{i1} > \delta_{i2} > \delta_{i3} \approx 0.  \label{deli}
\end{equation}

\noindent From microscopic reversibility or the principle of detailed balance, the product of the transition rates in the
clockwise and anticlockwise directions are equal \cite{hille},  and we may write

\begin{eqnarray}
\delta_{21} \frac{\alpha_{B1}}{\beta_{B1}}  & = &  \frac{\alpha_{A1}}{\beta_{A1}} = \frac{\alpha_{C}}{\beta_{C}}   \nonumber \\
\delta_{32} \frac{\alpha_{B2}}{\beta_{B2}}  & = &  \frac{\alpha_{A2}}{\beta_{A2}} = \frac{\alpha_{O}}{\beta_{O}}, \label{micro1}
\end{eqnarray}
	
\noindent where $\delta_{21} = \delta_{i2}/\delta_{i1} < 1$  and $\delta_{32} = \delta_{i3}/\delta_{i2} < 1$.

Assuming that $\beta_{i1} + \gamma_{i1} \gg \alpha_{A1}$, $\beta_{i2} + \gamma_{i2} \gg \alpha_{A2} + \beta_{A1}$
, $\beta_{i3} + \gamma_{i3} \gg \beta_{A2}$, and the first forward and  backward transitions are rate limiting
 ( $\beta_{ik} \gg \delta_{ik}$ and  $\gamma_{ik} \gg \alpha_{ik}$, for $k=1$ to $3$) \cite{vac3,lcfb},  the occupation 
probabilities $A_1$, $A_2$ and $A_3$ rapidly attain a quasi steady state
\begin{eqnarray}
A_1  & \approx &  \frac{\alpha_{i1} C_1 + \delta_{i1} B_1}{\beta_{i1} + \gamma_{i1} },  \\
A_2  & \approx & \frac{\alpha_{i2} C_2 + \delta_{i2} B_2}{\beta_{i2} + \gamma_{i2} }, \\
A_3  & \approx & \frac{\alpha_{i3} O + \delta_{i3} B_3}{\beta_{i3} + \gamma_{i3} }, 
\end{eqnarray}
	
\noindent and therefore, Eqs. (\ref{9c1}) to (\ref{9b3})  may be reduced to a six state master equation (see Fig. 2)
\begin{eqnarray}
\frac{dC_1}{dt} & = &  -(\rho_1 + \alpha_C)C_1(t) + \beta_C C_2(t) + \sigma_1 B_1(t) \label{6c1} \\
\frac{dC_2}{dt} & = &  \alpha_C C_1(t) - (\alpha_O + \beta_C + \rho_2) C_2(t) + \beta_O O(t) + \sigma_2 B_2(t) \label{6c2} \\
\frac{dO}{dt}   & = &   \alpha_O C_2(t) - (\beta_O + \rho_3)O(t) +\sigma_3 B_3(t)  \label{6o} \\
\frac{dB_1}{dt} & = &  \rho_1 C_1(t) -(\alpha_{B1} + \sigma_1) B_1(t)+\beta_{B1} B_2(t)  \label{6b1} \\
\frac{dB_2}{dt} & = &  \rho_2 C_2(t) + \alpha_{B1}B_1(t)  -(\alpha_{B2} +\beta_{B1} + \sigma_2) B_2(t) + \beta_{B2} B_3(t)  \label{6b2} \\
\frac{dB_3}{dt} & = &  \rho_3 O(t) + \alpha_{B2}  B_2(t) -(\beta_{B2} + \sigma_3) B_3(t), \label{6b3} 
\end{eqnarray}
where the derived forward and backward rate functions for  inactivation $\rho_k$ and $\sigma_k$ are, in general, voltage dependent
 \cite{cgh,ch,sh}
\begin{eqnarray}
\rho_k    & \approx &  \frac{\alpha_{ik} \gamma_{ik}}{\beta_{ik} + \gamma_{ik}},  \label{rho1}  \\
\sigma_k  & \approx & \frac{\delta_{ik} \beta_{ik} }{\beta_{ik} + \gamma_{ik}}.  \label{sig1}
\end{eqnarray}

If the conditions  $\beta_{ik} \gg \delta_{ik}$ and  $\gamma_{ik} \gg \alpha_{ik}$ for each $k$ are not satisfied, the inactivation of the 
Na+ current during a depolarizing potential may be bi-exponential.     From the assumptions of Eqs.
 (\ref{alfi}) and (\ref{deli}), the inactivation rate is not state dependent ($\rho_k = \rho_1$ for each 
$k$), the deinactivation rates $\sigma_1 > \sigma_2 > \sigma_2 \approx 0$, and therefore, from the 
microscopic reversibility conditions in  Eq.  (\ref{micro1}) 
\begin{eqnarray}
\beta_{B1} \alpha_C   & < &  \beta_{C} \alpha_{B1},  \label{mr1}  \\
\beta_{B2} \alpha_O   & < &  \beta_{O} \alpha_{B2}.  \label{mr2}
\end{eqnarray}

The nonzero eigenvalues of the characteristic equation for  Eqs. (\ref{6c1}) to (\ref{6b3}) are $\lambda_j = - \omega_j$
for $j = 1$ to 5 where $\omega_j$ may be approximated by the roots $\omega_{kF}$ and $\omega_{kG}$ of the 
cubic polynomials $F(\omega)$ and  $G(\omega)$ (see Fig. 3)  
 where  $\omega_{1F} < \omega_{2F} < \omega_{3F}$, $\omega_{1G} < \omega_{2G} < \omega_{3G}$,
\begin{eqnarray}
F(\omega)  & = &  \omega^3 - a_1 \omega^2  + a_2 \omega  - a_3,   \label{char2}
\end{eqnarray}
\begin{eqnarray}
a_1 & = & \alpha_O + \beta_O + \alpha_C + \beta_C +   \rho_1  +   \rho_2 +  \rho_3   \nonumber  \\
a_2 & = &  (\beta_C + \rho_2)(\beta_O + \rho_3) + \rho_3 \alpha_O  +  \nonumber \\
    &   &  (\alpha_C + \rho_1)(\alpha_O + \beta_O + \beta_C + \rho_2 + \rho_3) - \alpha_C \beta_C   \nonumber \\     
a_3 & = &  \alpha_O \rho_3 (\alpha_C + \rho_1) + (\beta_O + \rho_3)(\rho_2 \alpha_C + \rho_1 \beta_C + \rho_1 \rho_2 ),
  \label{aa}
\end{eqnarray}
and, in this section, it is assumed that $\rho_k = \rho_1$ for each $k$,
\begin{eqnarray}
G(\omega)  & = &  \omega^3 - b_1 \omega^2  + b_2 \omega  - b_3   \label{char3}
\end{eqnarray}
\begin{eqnarray}
b_1 & = &  \alpha_{B1} + \beta_{B1} + \alpha_{B2} + \beta_{B2} + \sigma_1  \nonumber \\
b_2 & = & \alpha_{B1} (\alpha_{B2} + \beta_{B2}) + \beta_{B1}\beta_{B2} + \sigma_1 (\beta_{B1} + \alpha_{B2} +  \beta_{B2}) \nonumber \\
b_3  & = &  \sigma_1 \beta_{B1} \beta_{B2}.  \label{bb}
\end{eqnarray}
Therefore,  for a depolarizing potential, we may define  $\omega_k  \approx \omega_{kF}$, 
$\omega_{k+2}  \approx \omega_{kG}$,  for $k=2,3$ whereas for a hyperpolarizing potential, 
 $\omega_k  \approx \omega_{kG}$, $\omega_{k+2}  \approx \omega_{kF}$. If   $\beta_h$ is  the rate of inactivation
and $\alpha_h$ is the rate of recovery from inactivation,   it may be shown from the
 characteristic equation that $\omega_{1} = \alpha_{h} + \beta_{h}   \approx \omega_{1G} + \omega_{1F}$ where
\begin{eqnarray}
\omega_{1G} & = &  \frac{b_3}{\omega_{2G}\omega_{3G}},   \label{om1g} \\
 \omega_{1F} & = & \frac{a_3}{\omega_{2F}\omega_{3F}}.   \label{om1f}
\end{eqnarray}

If $\alpha_{i1},\beta_{i1}, \gamma_{i1}$ are, in general, exponential functions of  $V$, the rate of inactivation
\begin{equation}	
\beta_h  \approx  \omega_{1F} =  \rho_1 = \frac{\alpha_{i1} }{1 + \beta_{i1} / \gamma_{i1} }    \label{ratebetah}
\end{equation}
has an exponential voltage dependence for small  clamp potentials but saturates for a larger depolarization when
$\alpha_{i1}$ is weakly dependent on voltage (see Fig. 4) \cite{hh}. It is assumed that the activation
sensors are independent and hence $\alpha_C = 2 \alpha_m, \alpha_O =  \alpha_m,  \beta_C =  \beta_m, \beta_O =  2 \beta_m$
\cite{arm} where  $\alpha_m$ and  $\beta_m$ are HH rate functions for Na+ channel activation, and may be approximated by
 two-stage expressions  (see Fig. 4) \cite{vac3}. If the DIII S4 sensor is
the slowest to deactivate  ($\beta_{B1} \ll \beta_{B2}$)  \cite{crgfb,shk},  $ \omega_{1G}$ and  $ \omega_{2G}$ are solutions of
 the equation
\begin{equation}	
\omega^2 - \omega (\alpha_{B1} + \beta_{B1} + \sigma_1) +  \sigma_1 \beta_{B1}  = 0, \label{charh}
\end{equation}
and the rate of recovery from inactivation  $\alpha_h \approx \omega_{1G}$.
For the rate functions of Fig. 4,  $\alpha_h \approx \sigma_1$ when $ \beta_{B1} \gg \sigma_1$ whereas
 for  $ \beta_{B1} \ll \sigma_1$ the rate of recovery for inactivation $\alpha_h \approx \beta_{B1}$.
 From the microscopic reversibility conditions of Eq. (\ref{micro1}), we may assume that 
$ \beta_{B1} \propto  \beta_C$ and $\alpha_{B1} \propto \alpha_{C}$ and therefore, $\alpha_h(V)$
  and $\beta_m(V)$ have a similar  voltage dependence for small hyperpolarizing potentials, which is consistent with
 the HH determination of the rate functions  ($\beta_m(V) \approx 57 \alpha_h(V)$)  \cite{hh}.

If the Na+ channel is depolarized to a clamp potential $V$  from a large hyperpolarizing holding potential,
 the solution of Eqs. (\ref{6c1}) to (\ref{6b3}) for $\sigma_1 > \sigma_2 > \sigma_3 \approx 0$ may be approximated by
 the solution of a master equation for which $\sigma_1 > \sigma_2,\sigma_3 = 0$
\begin{eqnarray}
C_1(t) & = &   k_1 C_{1s}   + \Sigma_{j = 1}^{5} k_{j+1} C_{1j} \exp (-\omega_{j}t)  \label{sl1} \\
C_2(t) & = &    k_1 C_{2s} + \Sigma_{j = 1}^{5} k_{j+1} C_{2j} \exp (-\omega_{j}t)  \label{sl2} \\
O(t)   & = &    k_1 O_{s}  + \Sigma_{j = 1}^{5} k_{j+1} O_{j} \exp (-\omega_{j}t)  \label{sl3} \\
B_1(t) & = &     k_1 B_{1s} + \Sigma_{j = 1}^{5} k_{j+1} B_{1j} \exp (-\omega_{j}t)   \label{sl4} \\
B_2(t) & = &     k_1 B_{2s} + \Sigma_{j = 1}^{5} k_{j+1} B_{2j} \exp (-\omega_{j}t)  \label{sl5} \\
B_3(t) & = &     k_1 B_{3s} + \Sigma_{j = 1}^{5} k_{j+1} B_{3j} \exp (-\omega_{j}t) , \label{sl6} 
\end{eqnarray}
where  
\begin{eqnarray}
k_1^{-1} & = &  \Sigma_{j = 1}^{2}( C_{js} + B_{js}) + O_s + B_{3s}, \label{kss}
\end{eqnarray}
\begin{eqnarray}
C_{1s} & = &   \sigma_1 \beta_{B1} \beta_{B2} E_0  \nonumber   \\
C_{2s} & = &   \sigma_1 \beta_{B1} \beta_{B2} \alpha_C (\beta_O +  \rho_1)  \nonumber  \\
O_{s}   & = &   \sigma_1 \beta_{B1} \beta_{B2} \alpha_C \alpha_O  \nonumber  \\
B_{1s} & = &     \beta_{B1} \beta_{B2} a_3   \nonumber \\
B_{2s} & = &     \beta_{B2} (\alpha_{B1} + \sigma_1) a_3 -   \rho_1 \sigma_1 \beta_{B2} E_0 \nonumber \\
B_{3s} & = &    \alpha_{B2} (\alpha_{B1} + \sigma_1) a_3 -    \rho_1 \sigma_1 \alpha_{B2} E_0 +  \nonumber \\
           &   &    \rho_1 \sigma_1  \alpha_C \alpha_O \beta_{B1}  \label{stat} ,
\end{eqnarray}
$E_0  = \rho_1 \alpha_O + (\beta_O + \rho_1)(\beta_C + \rho_1)$, $a_3$ is defined in Eq. (\ref{aa}), the amplitudes of the terms  for each state are dependent on
\begin{eqnarray}
C_{1j} & = &  E_{2}(\omega_{j})  \label{c1j} \\
C_{2j} & = &  - \alpha_C (\omega_{j} -  \beta_O - \rho_1 )  \label{c2j} \\
O_{j}   & = &  \alpha_C \alpha_O \label{oj} \\
B_{1j} & = &  -\frac{F(\omega_{j})}{\sigma_1}  \label{b1j}  \\
B_{2j} & = &  \frac{1}{\sigma_1 \beta_{B1}} \left( - \rho_1 \sigma_1 E_{2}(\omega_{j}) + (\omega_{j} - \alpha_{B1} - \sigma_1)F(\omega_{j})  \right)  \label{b2j} \\
B_{3j} & = &   -\frac{ \rho_1 \alpha_C \alpha_O +  \alpha_{B2} B_{2j}}{\omega_{j} - \beta_{B2}},  \label{b3j} 
\end{eqnarray}
and 
\begin{equation}	
E_{2}(\omega) = \omega^2 - \omega (\alpha_{O} + \beta_{O} + \beta_{C} + 2\rho_1) +   \rho_1 \alpha_O + (\beta_O + \rho_1)(\beta_C + \rho_1).  \label{e2}
\end{equation}

Applying the initial conditions ($C_1(0) = 1$ and $C_2(0) = O(0) = B_1(0) = B_2(0) = B_3(0) = 0$), the parameters
 $k_i, i=2$ to $6$ may be determined from the solution in Eqs. (\ref{sl1}) to  (\ref{sl6}). For a depolarizing potential,
 $\omega_{4} \approx \omega_{2G}$, $\omega_{5} \approx \omega_{3G}$ and therefore, from Eqs. (\ref{char3}) and
  (\ref{b1j}) to  (\ref{b3j}) , assuming that $\omega_{2} \neq \omega_{4}$ and   $\omega_{3} \neq \omega_{5}$ for a coupled model  of 
Na+ channel activation  and inactivation,  $F(\omega_{2}),F(\omega_{3}) \ll F(\omega_{4}),F(\omega_{5})$ and
$|B_{k2}|, |B_{k3}| \ll |B_{k4}|, |B_{k5}|$ for each $k$. Therefore, to satisfy the initial conditions,
 $k_5,k_6 \approx 0$ and 
\begin{eqnarray}
k_2 & = &   \frac{1 - k_1 \sigma_1 \beta_{B1} \beta_{B2} \omega_2 \omega_3}{(\omega_{2}-\omega_{1})(\omega_{3}-\omega_{1})} \label{k2} \\
k_3 & = &   -\frac{1 - k_1 \sigma_1 \beta_{B1} \beta_{B2} \omega_1 \omega_3}{(\omega_{2}-\omega_{1})(\omega_{3}-\omega_{2})} \label{k3} \\
k_4  & = &   \frac{1 - k_1 \sigma_1 \beta_{B1} \beta_{B2} \omega_1 \omega_2}
{(\omega_{3}-\omega_{1})(\omega_{3}-\omega_{2})}.  \label{k4}
\end{eqnarray}
That is, each term of the open state probability in Eq. (\ref{sl3}) with eigenvalue $\lambda = -\omega_{4}$ or 
$ -\omega_{5}$ where $\omega_{4}$ or $\omega_{5}$  may be approximated by the roots  $\omega_{2G}, \omega_{3G}$ of the polynomial
 $G(\omega)$ has an amplitude close to zero.  If it is assumed that $\rho_1 = \rho_2 = \rho_3$ and  the two 
activation sensors are independent, the roots of  Eq.  (\ref{char2})  are $\omega_{1F} = \rho_1$,
  $\omega_{2F} = \alpha_m + \beta_m + \rho_1$, $\omega_{3F} = 2(\alpha_m + \beta_m) + \rho_1$
 (see Fig. 5) and 
\begin{eqnarray}
O(t)   & \approx &  m(t)^2 h(t) \label{open} \\
m(t) & = &     \frac{\alpha_m}{\alpha_m + \beta_m}(1 - \exp [-(\alpha_m + \beta_m)t])   \label{mm} \\
h(t)  & = &   \frac{\alpha_h + \beta_h \exp [-(\alpha_h + \beta_h) t]}{\alpha_h + \beta_h}.  \label{hh} 
\end{eqnarray}

Therefore, following the application of a voltage clamp, the solution of the master equation lies on an invariant manifold,
defined by the eigenvectors with eigenvalues that are determined by the roots of the polynomial $F(\omega)$. 
If the deinactivation rates $\sigma_1 > \sigma_2 > \sigma_3 \approx 0$ are chosen to satisfy microscopic reversibility,
 it may be shown from  the numerical solution of the master equation or  from a more general form of the solution of
  Eqs. (\ref{6c1}) to (\ref{6b3})   that $O(t) = m(t)^2 h(t)$ is still a good approximation. From Eq.  (\ref{sl6}), the
probability for the inactivated state $B_3(t)$ has an initial delay  that diminishes with increasing depolarization, and
 may be approximated by a bi-exponential function \cite{cgabc} (see Fig. 6).

Assuming that the time-dependence of the
 Na+ channel open state probability is described by the solution of a phenomenological master equation, 
as well as the HH expression $m(t)^2 h(t)$, the conditions for model reduction, 
 $F(\omega_{2}),F(\omega_{3}) \ll F(\omega_{4}),F(\omega_{5})$  for depolarizing potentials, 
provide constraints upon the choice of  empirical activation rate functions. If $\omega_{2} \approx \omega_{4}$ 
and $\omega_{3} \approx \omega_{5}$ for a weakly coupled model of Na+ channel activation and inactivation,
 these conditions are not satisfied and therefore, the terms with eigenvalues 
$- \omega_{4}$  and $- \omega_{5}$  have a nonzero amplitude and also contribute to   the time-dependence of $O(t)$.

When the Na+ channel is hyperpolarized to a clamp potential $V$  from a large depolarizing holding potential,
 the solution of Eqs. (\ref{6c1}) to (\ref{6b3}) for  $\sigma_1 > \sigma_2,\sigma_3 = 0$ is given by  Eqs. (\ref{sl1}) to  (\ref{sl6}) 
where  $k_1$ and the stationary solution are defined in  Eqs. (\ref{kss}) and  (\ref{stat}), and for $j=1$ to $5$
\begin{eqnarray}
C_{1j} & = &  \frac{\sigma_1 \beta_{B1} \beta_{B2} E_{2}(\omega_j) }{ F(\omega_j) }  \label{c1jh} \\
C_{2j} & = &  -\frac{\sigma_1 \beta_{B1} \beta_{B2} \alpha_C (\omega_{j} - \beta_{O})}{F(\omega_j) } \label{c2jh} \\
O_{j}   & = &  \frac{\sigma_1 \beta_{B1} \beta_{B2} \alpha_O \alpha_C }{F(\omega_j) } \label{ojh} \\
B_{1j} & = &   -\beta_{B1} \beta_{B2}    \label{b1jh} \\
B_{2j} & = &  \beta_{B2} \left(-\frac{\rho_1 \sigma_1 E_2(\omega_j)  }{F(\omega_j) } + \omega_{j} - \alpha_{B1} - \sigma_1 \right) \label{b2jh} \\
B_{3j} & = &  - \frac{\rho_1 \sigma_1 \alpha_C \alpha_O  \beta_{B1} \beta_{B2} +  \alpha_{B2} F(\omega_j) B_{2j} }
{F(\omega_j) (\omega_{j} - \beta_{B2}) } . \label{b3jh} 
\end{eqnarray}
and $E_{2}(\omega)$  is defined in Eq. (\ref{e2}).

For the nonzero eigenvalues,  $\lambda_j = -\omega_{j}$  for $j = 1$ to $5$, of the characteristic equation, 
$\omega_{1} = \alpha_{h} + \beta_{h}  \approx \omega_{kG} + \omega_{kF}$  and we may define
\begin{equation}	
\omega_{k} \approx \omega_{kG},\omega_{k+2} \approx \omega_{kF}    \label{eigenhyp}
\end{equation}
for $k = 2,3$. Applying the initial conditions ($C_1(0) = C_2(0) = O(0) = B_1(0) = B_2(0) = 0$ and  $B_3(0) = 1$),  and  assuming
that $\omega_{2} \neq \omega_{4}$ and   $\omega_{3} \neq \omega_{5}$, from Eqs.  (\ref{char2}),  (\ref{c2jh})  and (\ref{ojh}), 
$F(\omega_{4}),F(\omega_{5}) \ll F(\omega_{2}),F(\omega_{3}) $,  $|C_{24}| \gg |C_{22}|, |C_{23}|, |C_{25}|$ and 
 $|O_{4}|, |O_{5}| \gg |O_{2}|, |O_{3}|$. Therefore,
to satisfy the initial conditions, $k_5,k_6 \approx 0$ and 
\begin{eqnarray}
k_2 & = &  -\frac{1}{(\omega_{2}-\omega_{1})(\omega_{3}-\omega_{1})} \label{k2h} \\
k_3 & = &   \frac{1}{(\omega_{2}-\omega_{1})(\omega_{3}-\omega_{2})} \label{k3h} \\
k_4  & = &   -\frac{1}{(\omega_{3}-\omega_{1})(\omega_{3}-\omega_{2})}  \label{k4h} .
\end{eqnarray}

Assuming that $\alpha_C = 2 \alpha_m, \alpha_O =  \alpha_m,  \beta_C =  \beta_m, \beta_O =  2 \beta_m$, and
that the DIII S4 sensor is the slowest to deactivate  ($\beta_{B1} \ll \beta_{B2}$)  \cite{crgfb,shk},
from Eq. (\ref{sl1}), we may write
\begin{eqnarray}
C_1(t) & \approx & \left(\frac{\beta_m}{\alpha_m + \beta_m} \right)^2 X \nonumber \\
       &   & \left(1 - \exp (-\omega_{1} t) 
\left[1 + \frac{\omega_{1}(1 - \exp (-(\omega_{2} - \omega_{1}) t)}{\omega_{2} - \omega_{1}} \right] \right), 
 \label{c1ht}
\end{eqnarray}
where $\omega_{1}$ and $\omega_{2}$ are solutions of Eq. (\ref{charh}).
Therefore, the time course of the recovery  from inactivation is bi-exponential and in agreement with the kinetics 
determined from Nav1.4 channels  \cite{cgabc} (see Fig. 7), but for large negative potentials, 
$\omega_{2} \approx \beta_{B1} \gg \omega_{1} \approx \sigma_1$, and Eq.  (\ref{c1ht}) reduces to the HH expression
 $C_1(t) = (\beta_m /(\alpha_m + \beta_m))^2 [1 - \exp (-\omega_{1} t)]$. For a weakly coupled master equation,
 the conditions $F(\omega_{4}),F(\omega_{5}) \ll F(\omega_{2}),F(\omega_{3})$  are not satisfied
and therefore, the terms with eigenvalues $- \omega_{4}$  and $- \omega_{5}$   also contribute to  $C_1(t)$.  

When the Na+ channel conductance is regulated by the activation of three voltage sensors in the
 DI, DII and DIII domains, and coupled to a two-stage inactivation process where $\sigma_1 > 0$ and  
 $\sigma_k = 0$ for $k > 1$  (see Fig. 8),  it may be shown that during a
 depolarizing clamp potential, $O(t) \approx  m(t)^3 h(t)$ where m(t) and h(t) are defined in Eqs. (\ref{mm}) and (\ref{hh}),
 and $\alpha_h$ and $\beta_h$ are approximated by the  the smallest roots  of two quartic polynomials  and may be
 determined from Eqs. (\ref{ratebetah}) and (\ref{charh}) (see Fig. 9).  This description of the time-dependence of the
 Na+ current  is still a good approximation if it is assumed that there is a separate opening step which is
 more rapid than the activation of the S4  sensors \cite{kb}. Similarly, during a hyperpolarizing clamp potential, 
 if  $\beta_{B1} \ll \beta_{B2},\beta_{B3}$  \cite{crgfb,shk}, the time-dependence
 of the closed state probability is a more general form of Eq. (\ref{c1ht}) (see Fig. 10)
\begin{eqnarray}
C_1(t) & = & \left(\frac{\beta_m}{\alpha_m + \beta_m} \right)^3 X \nonumber \\
 &   & \left(1 - \exp (-\omega_{1} t) 
\left[1 + \frac{\omega_{1}(1 - \exp (-(\omega_{2} -\omega_{1}) t)}{\omega_{2} - \omega_{1}}\right] \right).  \label{c1ht2}
\end{eqnarray}

 {\bf MASTER EQUATION MODEL OF A Na+ CHANNEL WITH A STATE DEPENDENT INACTIVATION RATE}

In this section, we consider the effect of an increase in the inactivation rate as the S4 sensors activate 
 ($\rho_1 < \rho_2 <\rho_3$) \cite{kb, cgabc,ab}, on the time-dependence of  m(t) and h(t). If it is assumed that
 $\alpha_{ik} = \alpha_{i1},  \gamma_{ik} = \gamma_{i1}$ for each k  and the DIV S4 rate functions satisfy
\begin{equation}	
 \beta_{i1} >  \beta_{i2} >  \beta_{i3},    \label{betai1}
\end{equation}
the derived inactivation rate functions $\rho_k$ are dependent on the closed or open state. In order  to satisfy
 microscopic reversibility, we may write
\begin{eqnarray}
\delta_{21} \frac{\alpha_{B1}}{\beta_{B1}}  & = &  \frac{\alpha_{A1}}{\beta_{A1}} = \frac{\alpha_{C} \beta_{i1}}{\beta_{C}\beta_{i2}} \\
\delta_{32} \frac{\alpha_{B2}}{\beta_{B2}}  & = &  \frac{\alpha_{A2}}{\beta_{A2}} = \frac{\alpha_{O}\beta_{i2}}{\beta_{O}\beta_{i3}}, 
\label{micro2}
\end{eqnarray}
	
\noindent  and therefore, from Eq.  (\ref{deli}), the rate functions satisfy the inequalities  (\ref{mr1})  and  (\ref{mr2}).

The  eigenvalues of the characteristic equation may be approximated by the roots of the cubic 
polynomials $F(\omega)$   and  $G(\omega)$ (see Fig. 11), and assuming that the activation sensors
 are independent,  and  $\omega_{1F} = \Delta_{1F}$, $\omega_{2F} = \alpha_m + \beta_m + \Delta_{2F}$, 
$\omega_{3F} = 2(\alpha_m + \beta_m) + \Delta_{3F}$  are the roots of $F(\omega)$ in Eq.  (\ref{char2}),
 the rate of inactivation for a depolarizing potential is (see Fig. 12)
\begin{equation}	
\beta_h  \approx  \frac{\alpha_m^2 \rho_3 + 2 \alpha_m \beta_m \rho_2 + \beta_m^2 \rho_1}{(\alpha_m + \beta_m)^2}
   \label{ratebetah2}
\end{equation}
which reduces to Eq. (\ref{ratebetah}) when the inactivation rate is not dependent on the closed or open state.
Therefore, the voltage dependence of $\beta_h$ has contributions from the inactivation rate $ \rho_k$ for each $k$, as well
as the activation functions $ \alpha_m$ and $ \beta_m$. However, most of the voltage dependence derives from 
the inactivation rate, and this is supported by the  increase in the  time constant for inactivation in
the charge-neutralized mutant Na+ channel DIV-CN  \cite{cgabc}.

 For a hyperpolarizing potential, assuming that 
$\omega_{1G} = \Delta_{1G} \ll \omega_{2G} = \alpha_{B2}+ \beta_{B1} + \Delta_{2G} \ll \omega_{3G} = 
\alpha_{B1} + \beta_{B2}+ \Delta_{3G}$ are the roots of the polynomial $G(\omega)$  in Eq.  (\ref{char3}),
it may be shown that  the rate of recovery from inactivation is
\begin{eqnarray}
\alpha_h & \approx &   \frac{\sigma_1 \beta_{B1} \beta_{B2}}
{(\alpha_{B2}+ \beta_{B1} + \Delta_{2G})(\alpha_{B1} + \beta_{B2} - \Delta_{2G})}   \label{alphah} \\
\Delta_{2G}  & = &   \frac{D_1 - \sqrt{D_1^2 - 4D_2} }{2}  \label{k4h},
\end{eqnarray} 
where $D_1 = \alpha_{B1} + \beta_{B2} - \alpha_{B2} - \beta_{B1}$,  
$D_2 = \alpha_{B1}  \beta_{B2}- \alpha_{B1}\beta_{B1} - \alpha_{B2} \beta_{B2}- 
\Delta_{1G}(\alpha_{B1} +  \beta_{B2}) + \sigma_1 \beta_{B2}$. If $\alpha_{B1},\Delta_{2G} \ll \beta_{B2}$,
 Eq.  (\ref{alphah})  reduces to
\begin{eqnarray}
\alpha_h & \approx &  \frac{\sigma_1 \beta_{B1} }
{\alpha_{B2}+ \beta_{B1} + \Delta_{2G}}, \label{ratealphah2}  
\end{eqnarray} 
and may be approximated by an exponential function of $V$  when $\beta_{B1} \ll \alpha_{B2} +  \Delta_{2G}$ \cite{hh}, whereas
 for more negative potentials, there is a gradual increase of $\alpha_h$ towards the saturation value $\sigma_1$,
 in accord with the rate of recovery for inactivated Na+ channels in hippocampal neurons (see Fig. 12) \cite{kb}.
The deinactivation rate $ \sigma_1$ is only weakly voltage dependent for   hyperpolarizing potentials as  
$ \beta_{i1} \gg \gamma_{i1}$, and therefore, most of  the voltage dependence of $\alpha_h$ derives from 
 the activation and deactivation functions between  inactivated states. For  the charge-neutralized mutant 
Na+ channel DIV-CN, the voltage dependence of  $\beta_{i1}(V)$ is reduced so that 
$ \beta_{i1}(V) \ll  \gamma_{i1}$
and  $\sigma_1 \ll  \delta_{i1}$, but the voltage dependence of $\alpha_{B2}$ and  $\beta_{B1}$ are not
affected, and therefore, the expression $\alpha_h$  is  in accord with the data describing a slow recovery
from inactivation   \cite{cgabc}.
 
 The solution of  the master equation, Eqs. (\ref{6c1}) to (\ref{6b3}), for
$\sigma_1 > \sigma_2, \sigma_3 = 0$ and $\rho_1 < \rho_2 <\rho_3$, is 
given by Eqs. (\ref{sl1}) to (\ref{sl6})  where  the stationary solution is
\begin{eqnarray}
C_{1s} & = &   \sigma_1 \beta_{B1} \beta_{B2} E_0   \nonumber  \\
C_{2s} & = &  \sigma_1 \beta_{B1} \beta_{B2} \alpha_C (\beta_O +  \rho_3)   \nonumber  \\
O_{s}   & = &   \sigma_1 \beta_{B1} \beta_{B2} \alpha_C \alpha_O  \nonumber \\
B_{1s} & = &      \beta_{B1} \beta_{B2} a_3  \nonumber \\
B_{2s} & = &     \beta_{B2} (\alpha_{B1} + \sigma_1) a_3 - \rho_1 \sigma_1 \beta_{B2} E_0  \nonumber \\
B_{3s} & = &      \alpha_{B2} (\alpha_{B1} + \sigma_1) a_3  - \rho_1 \sigma_1 \alpha_{B2} E_0 +  \nonumber  \\
       &   &   \rho_3 \sigma_1 \beta_{B1} \alpha_C \alpha_O,  \label{ss2}
\end{eqnarray}
$E_0 =  E_2(0) = \rho_3 \alpha_O + (\beta_O + \rho_3)(\beta_C + \rho_2)$, $a_3$ is defined in Eq.  (\ref{aa}),  and  the amplitudes of the terms of each state are dependent on
\begin{eqnarray}
C_{1j} & = &  E_2(\omega_j)    \label{c1j2} \\
C_{2j} & = &   -\alpha_C (\omega_{j} -  \beta_O -\rho_3 )  \label{c2j2} \\
O_{j}   & = &  \alpha_C \alpha_O \label{oj2} \\
B_{1j} & = &  -\frac{F(\omega_{j})}{\sigma_1}  \label{b1j2}  \\
B_{2j} & = &  \frac{1}{\sigma_1 \beta_{B1}} \left( - \rho_1 \sigma_1 E_2(\omega_j) + (\omega_{j} - \alpha_{B1} - \sigma_1)F(\omega_{j})  \right)  \label{b2j2} \\
B_{3j} & = &   -\frac{ \rho_3 \alpha_C \alpha_O +  \alpha_{B2} B_{2j}}{\omega_{j} - \beta_{B2}}.  \label{b3j2} 
\end{eqnarray}
where $E_2(\omega)  =  \omega^2 -  (\alpha_O + \beta_O + \beta_C + \rho_2 + \rho_3) \omega + 
 (\beta_C + \rho_2)(\beta_O + \rho_3) + \alpha_O \rho_3$.

 From Eq. (\ref{sl3}), applying the initial conditions 
($C_1(0) = 1$ and $C_2(0) = O(0) = B_1(0) = B_2(0) = B_3(0) = 0$), it may be shown that
 the time-dependence of the Na+ channel open state probability $O(t) \approx  m(t)^2 h(t)$ 
during depolarization to a clamp potential $V$  (see Fig. 13), where h(t) is defined in  Eq.  (\ref{hh}),  and  the  activation
 variable is dependent on both activation  and inactivation rate functions
\begin{eqnarray}
m(t) & = &     \frac{\alpha_m}{\alpha_m + \beta_m + \Delta}(1 - \exp [-(\alpha_m + \beta_m + \Delta)t])  \\  \label{mm2}
 \Delta   & = &  \frac{\alpha_m (\rho_2 +  2\rho_3)  + \beta_m (2\rho_1 + \rho_2)}{\alpha_m + \beta_m} - 3 \beta_h. 
 \label{delta} 
\end{eqnarray}
The probability for entry into the inactivated state $B_3$ may be also be approximated by a bi-exponential function (see Fig. 14),
 and the time course of recovery from inactivation is given by Eq. (\ref{c1ht})  (see Fig. 15).

{\bf  CONCLUSION
}   

Hodgkin and Huxley described the time-dependence of the Na+ current in the squid giant axon membrane in terms of the 
expression $m(t)^3 h(t)$ where the activation variable m(t) and inactivation variable h(t) satisfy rate equations  \cite{hh}.
An alternative description of the Na+ current in nerve and muscle membranes is provided by a master equation for coupled 
channel activation and inactivation processes where the deinactivation rate to the open state is small, but the
rate to closed states increases as the activation sensors in the domains DI, DII and especially DIII, deactivate. This
model accounts for the small Na+ current during  repolarization of an inactivated channel, the saturation of
the rate of recovery  from inactivation for large hyperpolarized potentials and the delay in the time-course of the 
recovery from inactivation \cite{kb}.
  If it is further assumed that inactivation is a two-stage process, 
the model can account for the kinetics and voltage dependence of Na+ inactivation for wild-type and mutant
 channels \cite{cgabc}.

In this paper, we consider the coupling between two voltage sensors that regulate the  Na+ channel  conductance and a two-stage 
inactivation process, where the first forward and backward inactivation transitions of the DIV S4 sensor are rate-limiting, 
ensuring that the inactivation decay during a  depolarizing voltage clamp is exponential.
As the Na+ current following inactivation is close to zero until the S4 sensors of the DIII, and either DI or DII domains deactivate, 
we have  assumed that $\sigma_1 >  \sigma_2 > \sigma_3 \approx 0$, and from the analytical solution of the reduced six state
  master equation  for a depolarizing clamp when  the inactivation rate is uniform between states and 
$\sigma_1 >  \sigma_2, \sigma_3 = 0$, 
the slowest eigenvalue is determined by the inactivation rate $\rho_1$, which has an exponential
voltage dependence, but saturates for a large depolarizing potential \cite{hh}. For a hyperpolarizing clamp of the Na+ channel, the
 rate of recovery from inactivation is dependent on the deinactivation rate $\sigma_1$ to the first closed state, as well as the
 rate functions of the DIII S4 sensor between inactivated states.  The voltage dependence of  the derived rate functions  for 
inactivation and recovery from inactivation have a  similar form to empirical expressions  for Na+ channels in the  squid axon  
\cite{hh},  hippocampal neurons  \cite{kb} and Nav1.4 channels \cite{cgabc}.

For a hyperpolarizing clamp potential, as the deinactivation rate $\sigma_1 >  \sigma_2 > \sigma_3 \approx 0$, 
 it may be assumed that the deactivation rate functions between closed and open states are greater than those between 
inactivated states ($\beta_{O} > \beta_{B2}, \beta_{C} > \beta_{B1}$),  in order to satisfy microscopic reversibility.
  Therefore, the  closed state terms with eigenvalues of the characteristic equation that are determined by the roots
 of the polynomial $F(\omega)$  have an amplitude that are close to zero, and as the DIII S4 sensor is the slowest to
 deactivate ($\beta_{B1} \ll \beta_{B2}$)  \cite{crgfb,shk},  the time-dependence of the recovery from
inactivation is bi-exponential, and therefore, in agreement  with the kinetic data from Nav1.4 channels \cite{cgabc}.

For a depolarizing clamp potential of a Na+ channel, assuming that $\beta_{O} > \beta_{B2}$ and  $\beta_{C} > \beta_{B1}$, 
each term of the open state probability with eigenvalue  $\lambda = -\omega$ where  $\omega$
  approximates a root of the polynomial $G(\omega)$, also has an amplitude close to zero. 
A further simplification is possible when it is assumed that the activation sensors 
are independent ($\alpha_C = 2 \alpha_m, \alpha_O =  \alpha_m,  \beta_C =  \beta_m, \beta_O =  2 \beta_m$) 
 and it may be shown that the time-dependence of the open state $O(t) =  m(t)^2 h(t)$. In most nerve membrane
 Na+ channels, the activation of three voltage sensors  regulate the  Na+ channel  conductance, and by application
 of  similar constraints on  the  activation and deactivation rate functions for  inactivated  and closed states,  
 the time-dependence $m(t)^3 h(t)$ of the Na+ current may be derived from the solution to an eight state
 master equation for coupled activation and inactivation. For models of the Na+ channel where the inactivation rate
 from the closed or open states  increases as the S4 sensors activate, a more general form for the Hodgkin-Huxley 
expression for the open state probability  may be derived where m(t) and h(t) are dependent on both 
activation and inactivation processes.

\newpage 
%Figure 1
\begin{figure*}
\begin{center}
\includegraphics[width=0.7\textwidth]{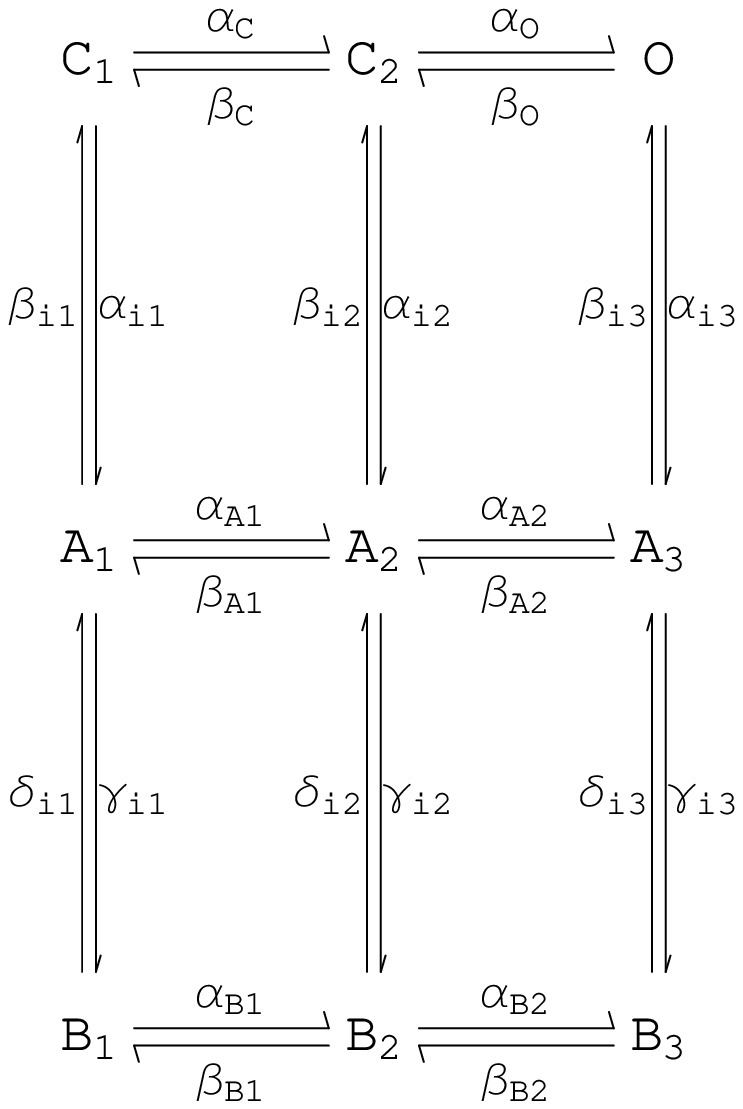}
\caption{
State diagram for Na+ channel gating where horizontal transitions represent the activation of two voltage sensors
 (DIII and either DI or DII)  that open the pore, and vertical transitions represent the two stage inactivation 
process of the DIV voltage sensor and the inactivation motif.
}
\end{center}
\end{figure*}

%Figure 2 
\begin{figure*}
\begin{center}
\includegraphics[width=0.7\textwidth]{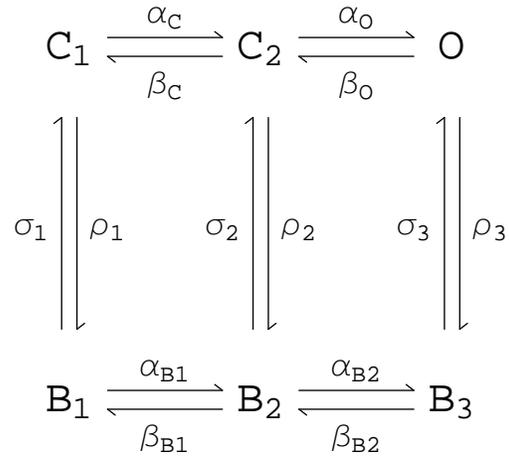}
\caption{
The nine state system for Na+ channel gating in Fig. 1 may be approximated by a six state system when 
$\beta_{ik} \gg \delta_{ik}$ and  $\gamma_{ik} \gg \alpha_{ik}$, for $k=1$ to $3$,
where $\rho_k$  and $\sigma_k$ are derived rate functions for a two-stage Na+ inactivation process, 
defined in Eqs. (\ref{rho1}) and (\ref{sig1}).
}
\end{center}
\end{figure*}

%Figure 3 
\begin{figure*}
\begin{center}
\includegraphics[width=0.7\textwidth]{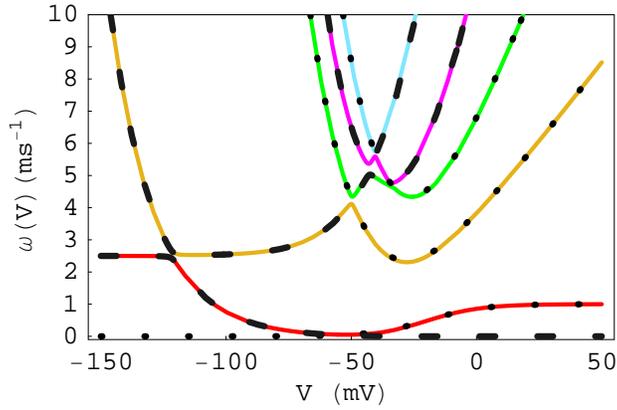}
\caption{
The voltage dependence of  $\omega_j = - \lambda_j$, $ j = 1$ to $5$, where $\lambda_j$ is a nonzero eigenvalue of
 the characteristic equation  of the master equation (solid line),  and the voltage dependence of $\omega_{jF}$  (dotted line) and
$\omega_{jG}$ (dashed line), the  roots of the cubic polynomials $F(\omega)$  and  $G(\omega)$,  where  the rate 
 functions are  $\alpha_{ik}(V) =  1$, $\gamma_{ik}(V) = \exp(3)$, $\beta_{ik}(V)=   \exp[-2(V-3)/25]$ for $k = 1$ to $3$,  
$\delta_{i1}(V)= 2.5$, $\delta_{i2}(V)= \delta_{i3}(V) = 0$, $\alpha_{C}  = 2 \alpha_{m } = \alpha_{B1}/3$,
$\alpha_{O}  = \alpha_{m } = \alpha_{B2}/3$, $\beta_C = \beta_m = 83.3 \beta_{B1}$,
$\beta_O = 2 \beta_m = 10 \beta_{B2}$ (ms$^{-1}$)  and $\alpha_{m }  = 0.1(V+25)/(1 - \exp[-(V+25)/10])$ and 
$\beta_m = 4 \exp[-(V+50)/18]$ are the HH rate functions for Na+ channel activation, assuming the resting potential is $V = -50$  mV.
}
\end{center}
\end{figure*}

%Figure 4 
\begin{figure*}
\begin{center}
\includegraphics[width=0.7\textwidth]{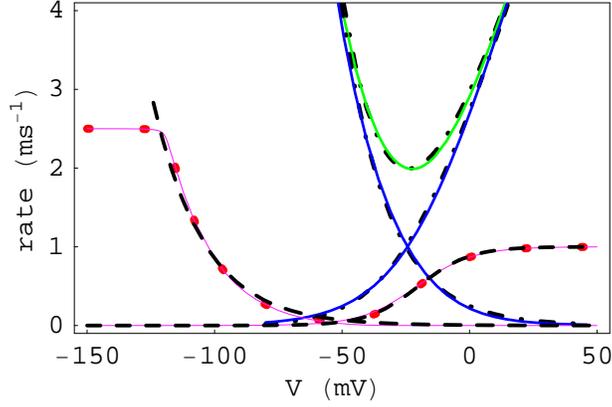}
\caption{
Voltage dependence of the HH Na+ channel inactivation rate functions $\beta_h = 1/(1+ \exp(-(20+V)/10))$ 
and $\alpha_h = 0.07 \exp(-(V+50)/20)$ (dashed line) may be approximated by the analytical
expression in  Eq. (\ref{ratebetah})  and by the smaller root of Eq.  (\ref{charh}) (solid line),
 derived from a master equation  for a six state system 
where activation and  two stage inactivation  are interdependent, and by the voltage dependence
of $\omega_1$  determined  numerically (dotted line) where the rate  functions are defined in Fig. 3. The HH Na+ channel
activation rate functions $\alpha_{m }  = 0.1(V+25)/(1 - \exp[-(V+25)/10])$, $\beta_m = 4 \exp[-(V+50)/18]$ and
$\alpha_{m} + \beta_m$  (dot dashed line)  may also be approximated by two stage expressions
$\alpha_{m,2}  = 2.3 \exp[0.32(V+50)/25](1 + 8.3 \exp[-1.3(V+50)/25])$ and 
$\beta_{m,2} = 4.2 \exp[-0.77(V+50)/18](1 + 8.3 \exp[-1.3(V+50)/25])$ (solid line)  [17].
}
\end{center}
\end{figure*}

%Figure 5 
\begin{figure*}
\begin{center}
\includegraphics[width=0.7\textwidth]{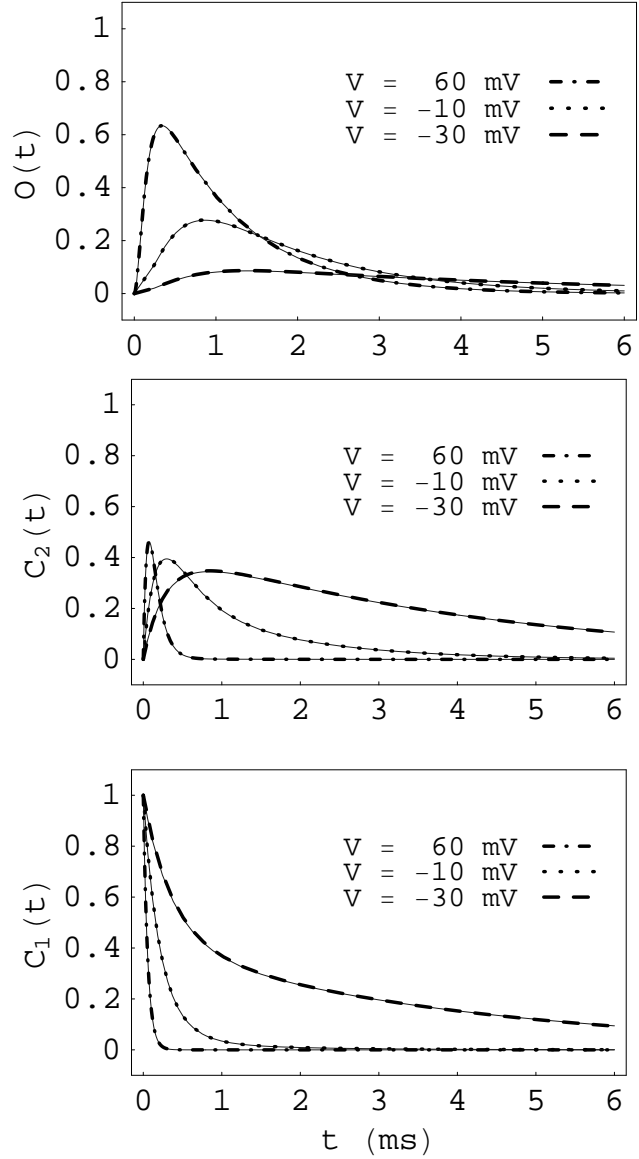}
\caption{
During a depolarizing clamp potential for the six-state system in Fig. 2 where activation and inactivation are interdependent, the open state probability  $O(t)$ (solid line) $\approx m(t)^2 h(t)$ (dashed, dotted or dot-dashed),  
$C_1(t)$ (solid line) $\approx [1 - m(t)]^2 h(t)$ (dashed, dotted or dot-dashed), $C_2(t)$ (solid line) $\approx 2m(t)[1 - m(t)] h(t)$ (dashed, dotted or dot-dashed), 
where m(t)  and h(t) are solutions of rate equations for activation and inactivation, and the rate functions are defined in Fig. 3.
}
\end{center}
\end{figure*}

%Figure 6 
\begin{figure*}
\begin{center}
\includegraphics[width=0.7 \textwidth]{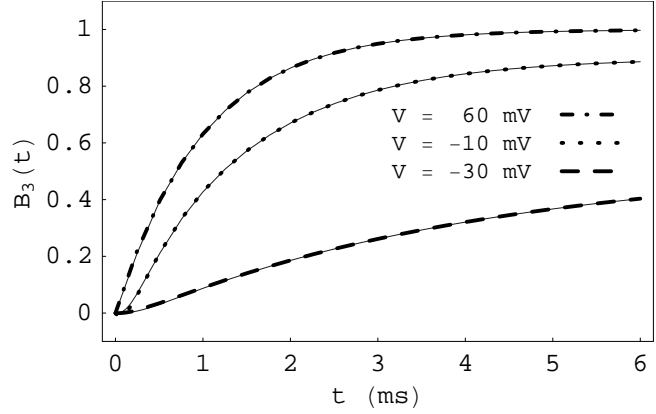}
\caption{
During a depolarizing clamp potential for a six-state system, the initial delay in the probability of the 
inactivated state $B_3(t)$ becomes less pronounced as the clamp potential increases, and may be approximated by
a bi-exponential for V = -30 mV and 60 mV, and by a tri-exponential for V = -10 mV (see Fig. 3).
}
\end{center}
\end{figure*}

%Figure 7 
\begin{figure*}
\begin{center}
\includegraphics[width=0.7 \textwidth]{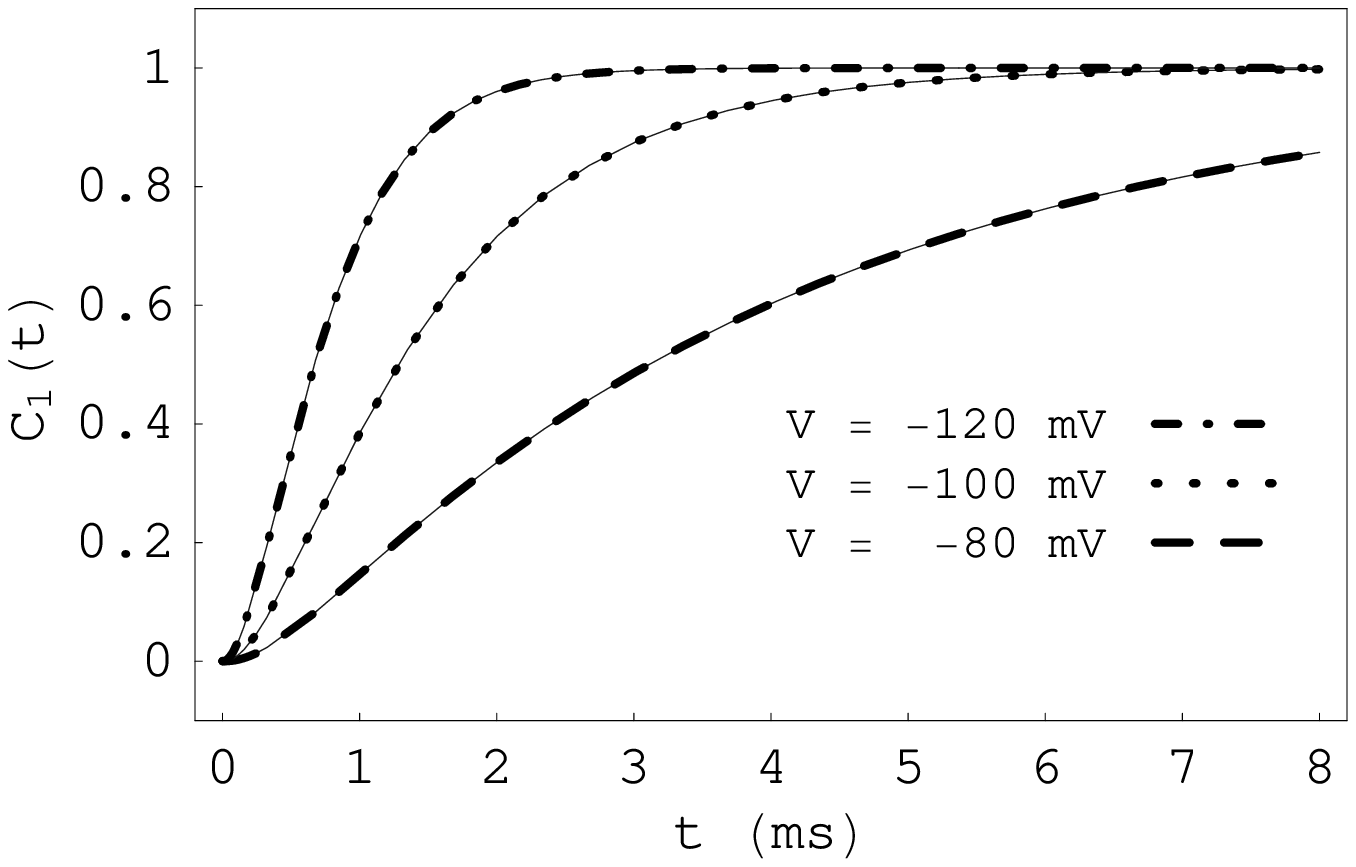}
\caption{
During a hyperpolarizing clamp potential for a six-state system, the first closed state  probability  $C_1(t)$ 
 may be described by the bi-exponential function in Eq. (\ref{c1ht}) (see Fig. 3).
}
\end{center}
\end{figure*}

%Figure 8 
\begin{figure*}
\begin{center}
\includegraphics[width=0.6 \textwidth]{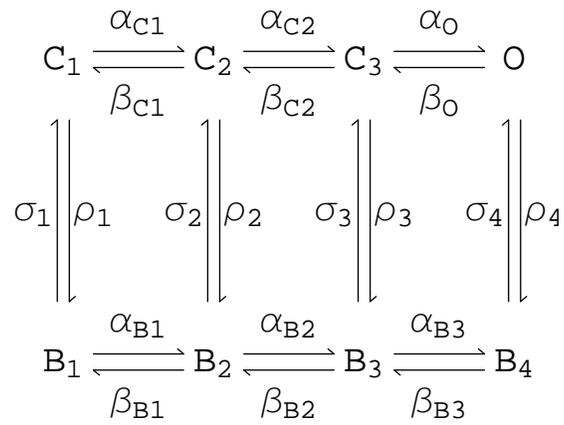}
\caption{
State diagram for Na+ channel gating where horizontal transitions represent the activation of  DI, DII and DIII voltage sensors
that open the pore,  and vertical transitions represent the derived rate functions for a two-stage Na+ inactivation process.
}
\end{center}
\end{figure*}

%Figure 9 
\begin{figure*}
\begin{center}
\includegraphics[width=0.7 \textwidth]{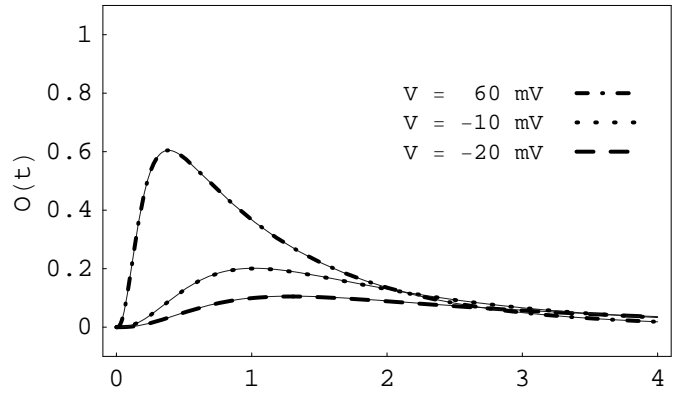}
\caption{
During a depolarizing clamp potential for the eight-state system in Fig. 8, the open state probability $O(t)$ (solid line) 
$\approx m(t)^3 h(t)$ (dashed, dotted or dot-dashed), where m(t)  and h(t) are solutions  of rate equations for activation and
inactivation, and the rate functions are $\alpha_{ik}(V) = 1$, $\gamma_{ik}(V) = \exp(3)$, $\beta_{ik}(V) = \exp[-2.3(V-13.2)/25]$
for k = 1 to 4, $\delta_{i1}(V)= 2.5$, $\delta_{ik}(V)= 0$ for k = 2 to 4, $\alpha_{m} = 0.1(V+25)/(1 - \exp[-(V+25)/10])$, 
$\beta_m = 4 \exp[-(V+50)/18]$,  $\alpha_{C1} = 3 \alpha_{m}$, $\beta_{C1} = \beta_m$, $\alpha_{C2} = 2 \alpha_{m}$, 
$\beta_{C2} = 2 \beta_m$,
 $\alpha_{O} = \alpha_{m}$, $\beta_{O} = 3 \beta_m$,  $ \alpha_{B1}= 3 \alpha_{C1}$, 
$\beta_{B1} = 0.012 \beta_{C1}$, $\alpha_{B2}= 3 \alpha_{C2}$, $\beta_{B2} = 0.8 \beta_{C2}$, $ \alpha_{B3}= 3 \alpha_{O}$, 
$\beta_{B3} = 0.8 \beta_{O}$ (ms$^{-1}$).
}
\end{center}
\end{figure*}

%Figure 10 
\begin{figure*}
\begin{center}
\includegraphics[width=0.7 \textwidth]{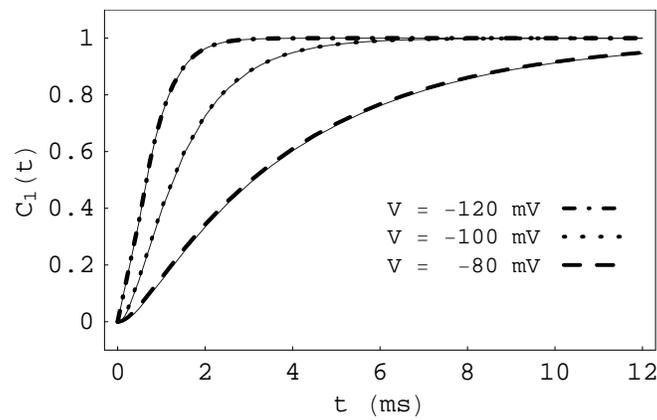}
\caption{
During a hyperpolarizing clamp potential for an eight state system, the first closed state probability $C_1(t)$  may be described by 
a bi-exponential function in Eq. (\ref{c1ht2}),  where the rate functions are defined in Fig. 9.
}
\end{center}
\end{figure*}

%Figure 11 
\begin{figure*}
\begin{center}
\includegraphics[width=0.7 \textwidth]{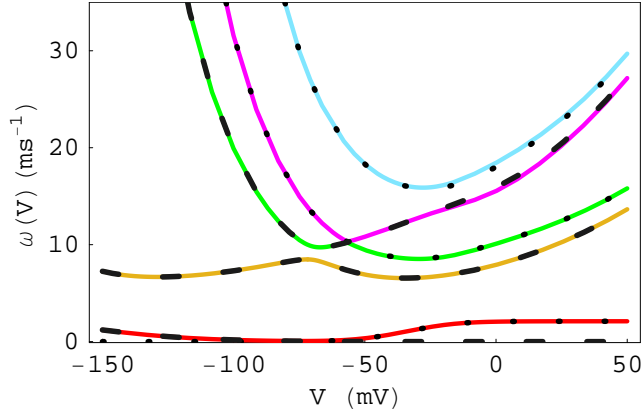}
\caption{
The voltage dependence of  $\omega_j = - \lambda_j$, $j = 1$ to $5$, where $\lambda_j$ is a 
nonzero eigenvalue of the characteristic equation of the master equation (solid line), and
the voltage dependence of $\omega_{jF}$ (dotted line) and $\omega_{jG}$ (dashed line), the
roots of the cubic polynomials $F(\omega)$  and  $G(\omega)$,  where rate functions are based
on those determined for Nav1.4 channels \cite{cgabc}, $\alpha_{ik}(V) = 2.1$, $\gamma_{ik}(V) = 24.9$, for 
$k = 1$ to $3$, $\beta_{i1}(V) = 12.8 \exp[-2.4V/25]$,
 $\beta_{i2}(V) = 1.6 \exp[-2.4V/25]$, 
$\beta_{i3}(V) = 0.2 \exp[-2.4V/25]$, $\delta_{i1}(V)= 2.5$, $\delta_{i2}(V)= \delta_{i3}(V) = 0$, 
$\alpha_{C} = 14.9 \exp[0.3V/25] = \alpha_{B1}$,
 $\alpha_{O} = 7.45 \exp[0.3V/25] = \alpha_{B2}$,
$\beta_C = 0.8 \exp[-0.9V/25] = 50 \beta_{B1}$ and
 $\beta_O = 1.6 \exp[-0.9V/25] = 3.3 \beta_{B2}$(ms$^{-1}$).
}
\end{center}
\end{figure*}

%Figure 12 
\begin{figure*}
\begin{center}
\includegraphics[width=0.7 \textwidth]{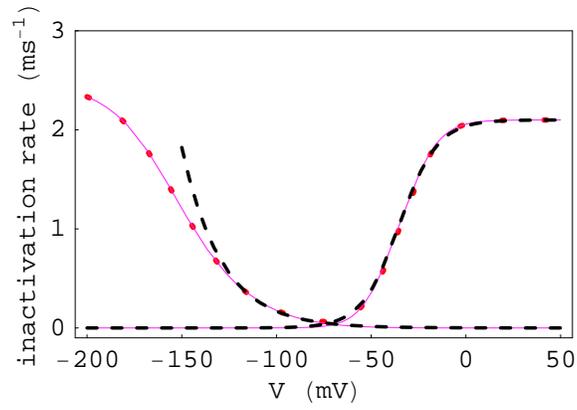}
\caption{
The voltage dependence of $\omega_1$, where $\lambda = -\omega_1$  is the slowest eigenvalue (dotted line)
 of the solution of a master equation for a six state
system
 with state dependent inactivation, may be approximated by the rate functions $\beta_h = 2.1/(1+ \exp(-(V+35)/10))$ and $\alpha_h = 0.065 \exp(-(V+80)/21)$
(dashed line) that have a similar voltage dependence to the  HH Na+ channel inactivation rate functions for the squid axon [1], and by the
 expressions $\beta_h$ and  $\alpha_h $  in Eqs.  (\ref{ratebetah2}) and  (\ref{ratealphah2}) (solid line) where the rate  functions are defined in Fig. 11.  
}
\end{center}
\end{figure*}

%Figure 13 
\begin{figure*}
\begin{center}
\includegraphics[width=0.7 \textwidth]{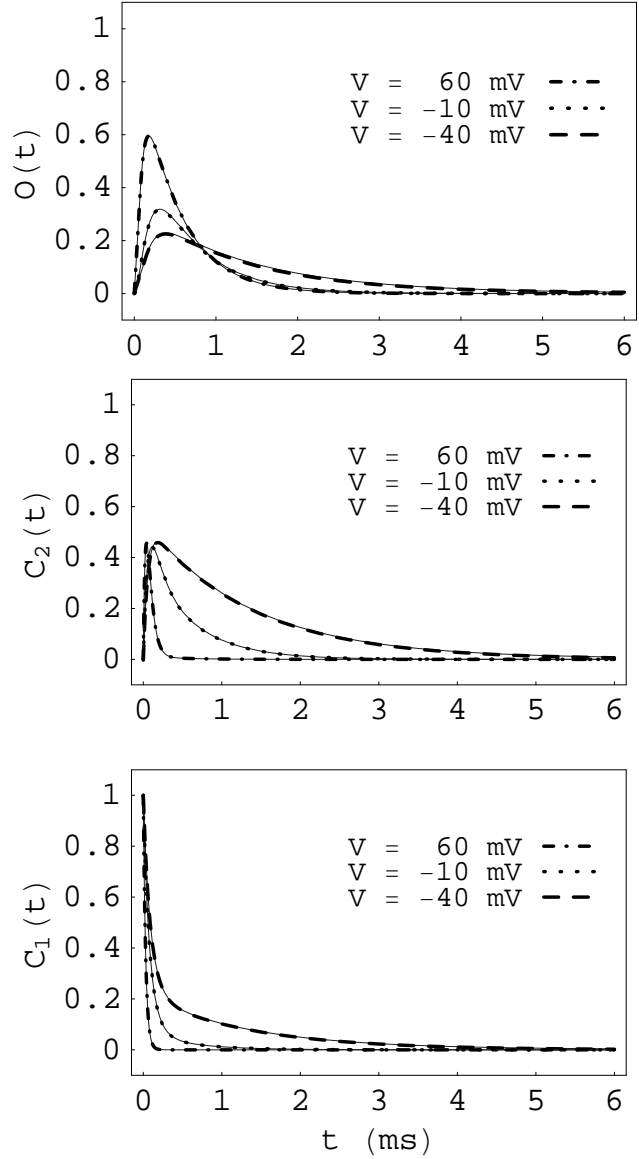}
\caption{
During a depolarizing clamp potential for a six-state system with state dependent inactivation,
 the open state probability $O(t)$ (solid line) $\approx m(t)^2 h(t)$ (dashed, dotted or dot-dashed),
the closed state  probabilities  $C_1(t)$ (solid line) $\approx [1 - m(t)]^2 h(t)$ (dashed, dotted or dot-dashed), $C_2(t)$ (solid line) $\approx 2m(t)[1 - m(t)] h(t)$ (dashed, dotted or dot-dashed), 
where the rate functions
 are defined in Fig. 11.
}
\end{center}
\end{figure*}

%Figure 14 
\begin{figure*}
\begin{center}
\includegraphics[width=0.7 \textwidth]{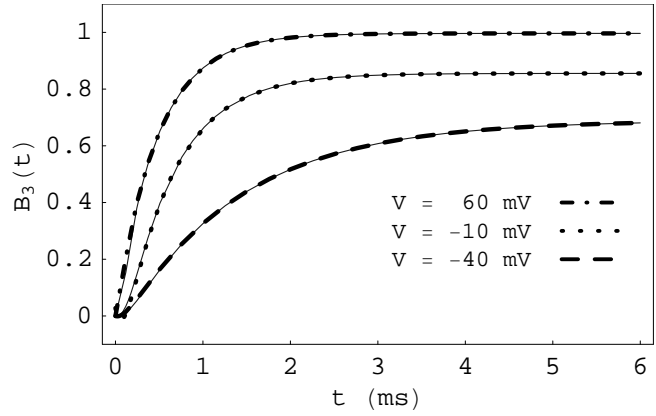}
\caption{
During a depolarizing clamp potential for a six-state system with state dependent inactivation, the initial delay in the probability of the inactivated state $B_3(t)$ 
becomes less pronounced as the clamp potential increases (see Fig. 11).
}
\end{center}
\end{figure*}

%Figure 15 
\begin{figure*}
\begin{center}
\includegraphics[width=0.7 \textwidth]{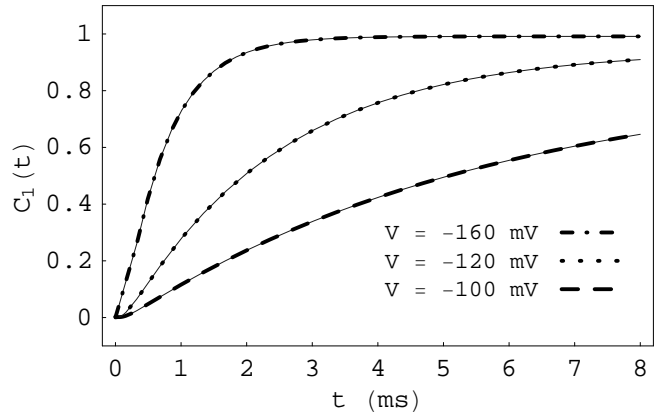}
\caption{
During a hyperpolarizing clamp potential for a six-state system with state dependent inactivation, the first closed state probability $C_1(t)$ 
may be described by the bi-exponential function in Eq. (\ref{c1ht}) (see Fig. 11).
}
\end{center}
\end{figure*}

\end{document}